# Windowed Compressed Spectrum Sensing with Block sparsity

Huiguang Zhang[1], Baoguo Liu[2,*]

**Abstract**

Compressed Spectrum Sensing (CSS) is widely employed in spectral analysis due to its sampling efficiency. However, conventional CSS assumes a standard sparse spectrum, which is affected by Spectral Leakage (SL). Despite the widespread use of CSS, the impact of SL on its performance has not been systematically and thoroughly investigated. This study addresses this research gap by analyzing the Restricted Isometry Property (RIP) of windowed Gaussian measurement matrices and proposing a novel block-sparse CSS model.

We introduce the Edge Zeroing Coefficient (EZC) to evaluate SL suppression and RIP impact, and the Window Scaling Coefficient (WSC) to quantify the effect on RIP. Our research investigates the influence of Window Function (WF) on signal sparsity and measurement matrices, and presents a block-sparse CSS model that considers component frequency distribution, signal length, windowing, and noise floor. Based on subspace counting theory, we derive sample bound for our model. The findings demonstrate that while WFs reduce SL, excessively small EZC and WSC values can negatively affect RIP quality and cause numerical instability during signal reconstruction. This highlights the delicate balance required when applying WFs in CSS. Our block-sparse approach enables precise compression and reconstruction, particularly for high noise floor and super-sparse signals. This study provides a framework for optimizing CSS performance when dealing with SL and sparse signals, offering insights for improving signal reconstruction quality in various applications

# Introduction

## Motivation

In wireless communications[1], radar systems[2], medical imaging[3] and online mechanical fault detection[4], power quality monitoring[5]. Increasing broadband and high-throughput data pose significant challenges to traditional spectrum sensing systems. To address these challenges, various CSS technologies have been developed, such as Compressed Covariance Sensing (CCS) [6], Analog-to-Information Converter (AIC) [7] and Modulated Wideband Converter (MWC) [8]. Compared to traditional spectrum sensing systems that must satisfy the Nyquist sampling theorem, CSS technologies can accurately sample, transmit and store signals in a more "economical" manner[9]. This is mainly due to the fact that CSS takes full advantage of the prior knowledge that signals are sparse in the frequency domain[10].

However, In practice, the sparsity of a signal in CSS is often unavoidably affected by factors such as SL, noise, and basis mismatch. SL causes signal energy to spread into nearby intervals of real frequency points, transforming the distribution of non-zero elements from discrete point-like to banded. This phenomenon shifts the signal's sparsity from standard to block sparsity, a concept introduced by Eldar [11]and Lu et al[12]. This seriously affects the sampling efficiency and reconstruction accuracy of CSS, posing significant challenges to the practical application of CSS.

Nevertheless, despite that some researchers have paid attention to the effects of SL on CSS and have conducted a number of studies on this topic. For instance, Chi et al. Note that physical signals in the DFT or predefined bases are rarely sparse due to mismatches between assumed and actual sparsity bases, leading to significant recovery errors[13]. Duarte et al. Propose a structured sparse signal model using an oversampled DFT framework to address SL in compressed sampling, develop the Spectral Iterative Hard Thresholding (SIHT) algorithm, and investigate interpolation methods for improved frequency resolution[14]. Zhong et al. Introduce a method that uses Blackman WF and compressed sampling and selects the most appropriate WF to reduce SL, increase signal sparsity, and improve the accuracy of ultra-high harmonic signal reconstruction[15]. There is still a lack of systematic research on the application of WFs in CSS and the construction of CSS models based on block sparse structures.

In contrast to the limited research in CSS, the impact of SL has been extensively studied in conventional spectrum analysis. Various effective methods have been developed to mitigate this impact, including extending the sample length[16], employing WF [17], and utilizing interpolation algorithms [18]. While these conventional techniques successfully mitigate SL, they face constraints in real-world CSS implementations. Prolonging the sampling duration can diminish SL effects, but this approach may not be viable in rapidly evolving environments or real-time monitoring systems due to device limitations [19].The successful implementation of interpolation algorithms requires accurate extraction of the spectrum peak and its neighboring points' values [20], but this task is challenging in conventional sparse recovery contexts. For instance, the interpolated Discrete Fourier Transform (IpDFT) method can effectively mitigate the impact of sampling on the continuous spectrum by interpolating the spectrum samples suitably in the neighborhood of each spectral peak [20]. However, the accuracy of these algorithms can be compromised by inherent noise and the non-uniformity of the sampling process [19], WFs offer a promising alternative, as they can be integrated into the measurement matrix through direct coupling with the Fourier transform matrix [22]. This approach circumvents the limitations associated with interpolation algorithms and provides a more flexible solution for addressing SL in CSS applications. The use of windowed interpolation FFT has been shown to suppress the effects of spectrum leakage and improve estimation accuracy, particularly in environments where synchronization of signal sampling is difficult [21]. Thus, the combination of these methods can enhance the robustness of spectral analysis in CSS.

In summary, SL severely affects the sampling efficiency and reconstruction accuracy of CSS, posing significant challenges to its practical application. Although a limited number of studies have investigated the effect of SL on CSS, there remain research gaps in the following areas:

1. **Minimizing SL with WF in CSS**: Implement effective measures such as WF to control the number of non-zero elements above the threshold. This approach necessitates a detailed study of how these measures affect the random projection behavior of the measurement matrix.

2. **Leveraging Block Sparse Structure**: Develop appropriate signal models and reconstruction algorithms that utilize the sparse block structure. This strategy aims to ensure lower sample bounds and higher reconstruction accuracy.

**Main contribution**

To fill the gap in WFs' application in CSS, we investigate the RIP property of windowed Gaussian measurement matrices based on the mechanism of WFs and Gauss 2-stablity. Additionally, we construct sample bound models for multi-frequency signals to circumvent the SL-related drawbacks in the conventional CSS. Below is a brief summary of our main contribution.

1) **Edge Zeroing Coefficient: A Comprehensive Evaluation Metric**

We introduce the Edge Zeroing Coefficient (EZC) as a novel metric for evaluating the effectiveness of windowing functions in compressed sensing systems. The EZC assesses WF performance from two critical perspectives:

1. **RIP stability:** It evaluates how the edge elements of the WF approaching zero affect the stability of the measured inverse matrix map, which is crucial for maintaining the RIP.

2. **Improving sparsity:** The EZC quantifies the WF's ability to SL and thereby enhance signal sparsity.

By combining these aspects, the EZC provides a comprehensive method for evaluating WF performance, offering researchers and engineers a valuable tool for optimizing WF selection and design in various signal processing scenarios.

2) **RIP model for windowed measurement matrices：**

Then, we propose a novel RIP model for windowed sub-Gaussian measurement matrices that exploits the 2-stability of Gaussian distributions. The main features and contributions of the model are as follows:

1. **WF integration:** The model considers the effect of a normalized WF on the RIP of standard sub-Gaussian random measurement matrices.

2. **Window Scaling Coefficient (WSC):** We introduce WSC to establish a correspondence between the RIP of standard random measurement matrices and that of windowed random matrices.

3. **Theoretical basis:** This approach provides a simple way to analyze the influence of WFs on sub-Gaussian random matrices in compressed sensing applications.

By integrating these elements, our model provides a comprehensive framework for understanding and optimizing the performance of window measurement matrices.

3) **Sample bound model of the measurements required for a windowed block CSS:**

Finally, This study introduces advancements in window block CSS, focusing on the following features:

   1. **Block Lower Sparse Bound Model**: We propose a novel model for predicting the minimum dimensions required in window block CSS. This model uniquely incorporates critical factors such as component frequency distribution, signal length, and noise floor, providing a more accurate estimation of the necessary dimensions for effective signal reconstruction.

   2. **Optimized WF Strategy**: We develop a comprehensive strategy for optimizing WF selection and utilization. This optimization significantly reduces the number of measurements required for signal reconstruction while enhancing the accuracy of CSS when processing sparse signals.

By integrating these research, our research provides a framework for advancing the field of compressed sensing, particularly in applications involving sparse signals and SL challenges. This integrated approach opens new avenues for more efficient and accurate signal processing techniques in various practical applications, potentially revolutionizing the way we approach CSS in complex signal environments.

**Organization**

The remainder of this paper is organized as follows. Section 2 provides the reader with the background knowledge necessary to understand the following sections, presenting a brief overview of the basic theory related to the properties of the RIP for WFs and block-sparse signals. Section 3 presents the core of our theory, deriving the mathematical model of block RIP for windowed sensing matrices based on compressive sensing, WF properties, and multivariate statistics. Furthermore, we propose the EZC to quantify the influence of different WFs on various facets of spectrum sensing, including block RIP. Finally, we present a sample bound in the context of block sparsity for windowed CSS. In the discussion section, we explain the importance of the research, summarize the limitations of our study, and provide insights into future research approaches.

**Preliminary Theories**

Before delving into the impact of WFs on CSS, we first review the basics of SL, WFs, and CSS theory. This review will provide a theoretical framework for the system so that readers can better understand our main work. This overview would provide the necessary theoretical framework to help readers better understand our main work.

**Spectrum Leakage and Window fucntion**

For spectrum analysis applications, Fourier analysis, represented by Fast Fourier Transform (FFT), is the most widely used due to its high stability and favorable computational efficiency. The Discrete Fourier Transform (DFT) can be expressed as follows:

$$X[k] = \sum_{n=0}^{N-1} x[n] \cdot e^{-\iota \frac{2\pi}{N} kn}, k = 0,1,\ldots,N-1 \tag{1}$$

Where Greek letter $\iota$ (iota) denotes complex unit , Note that it is different from the lowercase of L. Because we need to frequently use $i$ and $j$ to denote subscripts in subsequent proofs, we cannot use $i$ and $j$ to denote complex units here.

Theoretically, the Fourier transform can extract the component's accurate characteristic parameters, provided the Nyquist theorem is satisfied and the sampling time is infinitely long. However, in the real world, the sample length of the received signal is always limited and the frequency of the component signals may not be precisely known. The inevitable result of a finite signal length is: signal truncation and unsatisfactory spectral resolution. This leads to boundary discontinuity and mismatches in the Fourier analysis basis and ultimately to SL [23]. Therefore, SL has become a major challenge in practical spectral analysis[24]. The problem of SL is typically addressed with WFs and increasing the sample length, as discussed in detail below.

## WFs' impact on discontinuity

When performing the DFT, we implicitly assume that $x[n]$ is a period of an infinitely extended periodic signal $\tilde{x}[n]$ with period $N$:.

$$\tilde{x}[n] = x[n \bmod N], n = -\infty, \dots, -1, 0, 1, \dots, \infty \tag{2}$$

If the original signal $x[n]$ is not periodic within the duration $N$, a jump occurs at the boundaries when $\tilde{x}[n]$ is formed by infinitely repeating $x[n]$. These discontinuity can be represented as the sum of the original signal and a discontinuous signal $d[n]$:

$$\tilde{x}[n] = x[n] + d[n], n = -\infty, \dots, -1, 0, 1, \dots \tag{3}$$

The DFT of $\tilde{x}[n]$ is the sum of the DFTs of $x[n]$ and $d[n]$:

$$\tilde{X}[k] = X[k] + D[k], k = 0, 1, \dots, N-1 \tag{4}$$

The discontinuity signal $d[n]$ has a non-zero spectrum $D[k]$ that spreads across almost all frequencies, causing SL in the DFT of $\tilde{x}[n]$.

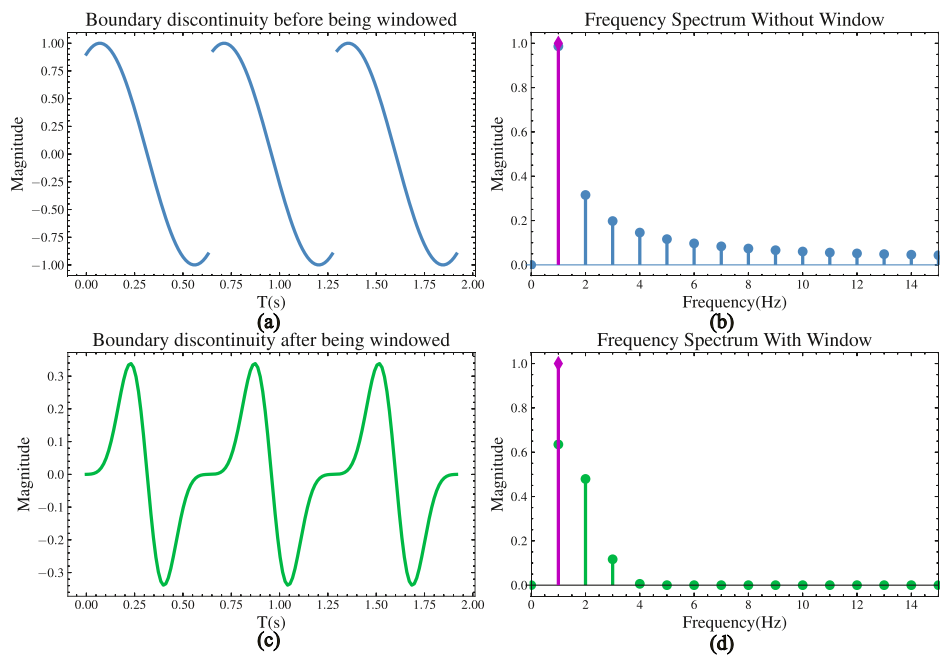

Figure 1 boundary discontinuity before and after windowed

The diagram above demonstrates the spectral distribution before and after windowing of a periodically discontinuous function.

Subfigures (a) and (b) show the result of the periodic extension of a finite-length signal and its spectral distribution. It is clearly evident that the discontinuity at the boundary of the time-domain signal leads to spectral spreading. Subfigures (c) and (d) illustrate the improvement of the discontinuity at the boundary after windowing and its impact on SL.

The WF mainly involve the size and size of the jump, including the amplitude of the jump at the boundary, the size of the transition from the boundary to the center, and the order of the continuous derivative. In addition, zeroing the WF at the boundary ensures the continuity of the boundary, and the advantage is that it is not necessary to estimate the boundary value of the original signal. The most commonly employed WFs are Hamming window, Hanning window, Blackman window, Kaiser window, and flat-top window. Each WF possesses distinct advantages and disadvantages with regard to spectral resolution, amplitude accuracy, and sidelobe rejection. The efficacy of a WF in mitigating SL is assessed through three conventional criteria: main lobe width, side lobe level, and side lobe attenuation rate. These three are evaluated based on the outcome after applying the WF, which makes quantification challenging. The above is an intuitive description of the role of WFs. In other words, by controlling these indicators, WFs with better effects can be constructed. However, there is still a lack of practical and universal quantitative construction standards, which is also part of the research content of this work.

**Signal length's and Basis mismatch**

In this subsection, we investigate the effect of fundamental mismatch caused by unsatisfactory sampling length on SL. SL due to basis mismatch occurs when the frequency components of a signal do not align perfectly with the discrete frequency grid used in the Discrete Fourier Transform (DFT) or Fast Fourier Transform (FFT).

Consider a single frequency component a continuous-time signal x(t) with:

$$x(t) = A exp(\iota\, 2\pi f_0 t + \phi_0) \tag{5}$$

where A is the amplitude and $f_0$ is the true frequency, $\phi_0$ is the phase.

We sample this signal at a rate fs for N points, giving us a discrete-time signal:

$$x[n] = A \exp\left(\frac{\iota\, 2\pi f_0 n}{f_s}\right), \quad n = 0, 1, \dots, N-1 \tag{6}$$

The DFT basis functions are:

$$F_k[n] = \exp\left(-\frac{\iota 2\pi k n}{N}\right), \ k, n = 0, 1, \ldots, N-1 \tag{7}$$

The DFT coefficient at frequency index k is:

$$\begin{aligned}X[k] &= \sum_{n=0}^{N-1} x[n] \cdot \psi_k[n] \\ &= A \cdot \sum_{n=0}^{N-1} \exp\left(\iota 2\pi \left(\frac{f_0}{f_s} - \frac{k}{N}\right) n\right)\end{aligned} \tag{8}$$

If $\frac{f_0}{f_s}$ is not an integer multiple of 1/N, we have basis mismatch. Let's define:

$$\delta = -\left\lfloor \frac{f_0}{f_s} \right\rfloor \tag{9}$$

where $\delta$ represents the mismatch.

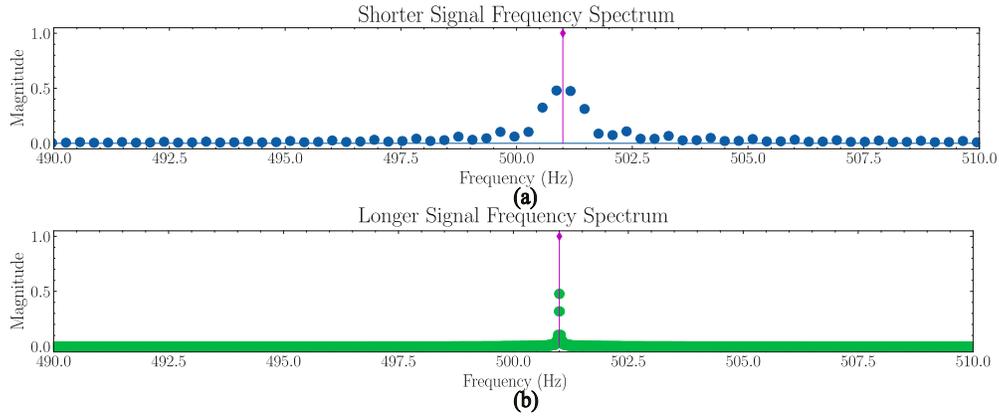

Figure 2 basis mismatch under long signal and short signal

In subfigure (a), the shorter signal demonstrates pronounced SL, with energy distributed over an extended range around the true frequency. This is primarily attributed to a considerable discrepancy between the signal's true frequency and the Fourier basis, which results in a significant mismatch issue. In comparison, subfigure (b) reveals that as the signal duration extends, the alignment between the signal frequency and the Fourier basis enhances. Consequently, the principal component predominantly overlaps with the true spectrum, enabling a more focused concentration of energy at the target frequency

The DFT coefficient can now be written as:

$$X[k] = A \cdot \sum_{n=0}^{N-1} \exp\left(\iota 2\pi \left(\delta + \frac{k'}{N}\right)n\right)$$
$$= A \cdot \frac{1 - \exp\left(\iota 2\pi \left(\delta + \frac{k'}{N}\right)N\right)}{1 - \exp\left(\iota 2\pi \left(\delta + \frac{k'}{N}\right)\right)} \tag{10}$$

Where $k' = k - \left[\left(\frac{f_0}{f_s} \cdot N\right)\right]$

This can be simplified to:

$$X[k] = A \cdot \exp\left(\iota \pi \left(\delta + \frac{k'}{N}\right)(N-1)\right) \cdot \frac{\sin\left(\pi \left(\delta + \frac{k'}{N}\right)N\right)}{\sin\left(\pi \left(\delta + \frac{k'}{N}\right)\right)} \tag{11}$$

The magnitude of $X[k]$ is:

$$|X[k]| = A \cdot \left|\frac{\sin(\pi \delta N)}{N \cdot \sin\left(\frac{\pi \delta}{N}\right)}\right| \tag{12}$$

(12) shows the SL effect. When $\delta \neq 0$ (i.e., there's a basis mismatch), the energy of the signal is spread across multiple frequency bins, rather than being concentrated in a single bin. This spreading of energy is what we call SL.

The magnitude of SL depends on $\delta$, which represents how far the true frequency is from the nearest DFT bin. The larger the mismatch ($\delta$), the more severe the SL.

Truncation and basis mismatch are two distinct phenomena, both of which can result in spectral loss, but differ significantly in their nature and effects. The truncation that occurs when a signal is observed over a finite time is analogous to applying an rectangular WF to the signal. This leads to global spectral diffusion, where the energy spreads across the entire frequency spectrum. Basis mismatch, on the other hand, is an intrinsic effect that occurs when the frequency components of the signal do not perfectly match the discrete frequency grid used in the analysis. This misalignment leads to local spectral diffusion, where energy escapes primarily into neighboring frequency ranges. The above discussion summarizes the principles of the WF and increasing the sample length to reduce SL. Despite these two methods, controlling SL remains a challenge and requires balancing the negative effects of the WF. Relatively speaking, the windowing function is a more economical means because it can achieve a more concentrated spectral distribution with fewer samples.

# Structured sparse Signal and Block RIP

**Compressed sensing**

A signal $x$ is sparse or compressible if its energy is concentrated in a limited number of K components. Candes and Tao's research demonstrates that a compressible signal $x$ can be accurately reconstructed from very limited measurements $y$ obtained via a measurement matrix $\Psi$. And the problem can be formulated an linear under-determined problem as follows:

$$y = \Psi x + n \quad (13)$$

For many signals, the original signal x is sparse only on a specific transformed domain, namely

$$\hat{x} = \Phi x \quad (14)$$

The sensing matrix and sparse transformation matrix are often combined into a single matrix called the measurement matrix for computational efficiency

$$\Theta = \Psi \Phi \quad (15)$$

And (13) can be represented as

$$y = \Psi \Phi x + n \quad (16)$$

Candes et al. further posits that to complete the aforementioned compressed sampling and reconstruction process, the measurement matrix utilized must satisfy a set of conditions, known as the RIP conditions. The RIP is a key concept in compressed sensing theory. A $M \times N$ matrix $\Psi$ is said to satisfy the RIP of order k if there exists a constant $\delta_k$ (called the RIP constant) such that for all k-sparse vectors x:

**Definition 1.** To definition of :

$$(1 - \delta_k)||x||_2^2 \leq ||\Psi x||_2^2 \leq (1 + \delta_k)||x||_2^2 \quad (17)$$

Where: $||x||_2^2$ denotes the $l_2$-norm (Euclidean norm) of the vector $x$. $\Psi$ denotes the measurement matrix. $\delta_k$ is the RIP constant, which is a small positive value, typically much less than 1.

From the random projection theory of high-dimensional polyhedron, we can explain RIP as follows, the RIP requires that the sensing matrix $\Psi$ **approximately** preserves the Euclidean distance of any k-sparse vector pairs . This is a crucial property for the success of sparse signal recovery algorithms.

Determining the RIP parameters for a given measurement matrix is generally an NP-hard problem for its exponential measurements. And there is no general closed method to calculate the restricted isometric constant (RIC). However, to the work of Candes and Tao[25], the random matrices eigenvalue theory has shown that some random matrices have a overwhelming probability of satisfying the RIP. These matrices include Gaussian matrices, Bernoulli matrices and partial Fourier matrices. Furthermore, Baraniuk et al. have provided a simple proof of RIP using the Johnson-Lindenstrauss (J-L) lemma[26], and Romberg examined the incoherence required for the measurement matrix[27]. These contributions have significantly advanced the development of compressed sensing theory and the understanding of sparse signals.

**Model-based Compressed sensing and Union of Subspace**

The conventional sparse models are based on a simplified signal model. This model only considers the number of non-zero elements, without incorporating prior knowledge such as the distribution pattern of non-zero elements. However, in many fields, including spectral sensing, DNA microarrays, magneto encephalo graphy(MEG), image processing, and communication systems, the distribution of non-zero elements in sparse signals often exhibits more complex structures that cannot be fully expressed by sparsity alone [28]. One situation is that in wavelet transforms, the determinant coefficients tend to form a cluster and connect into a root subtree[29]. Another situation is that the non-zero coefficients may appear in blocks [30]. Such limitations can be exploited to create a more concise model. For example, such a structure may represent the sparse combination of subspaces of finite dimension, where the signal contains components in k subspaces selected from n possibilities (i.e., the problem of selecting k from n) [31]. In contrast to compressive sensing (which relies on the signal being sparse and containing a small number of non-zero elements), the union subspace model also exploits the prioritization of the positions of the non-zero elements. This reduces the number of measurements even further.

To process such signals more effectively, researchers have proposed structural sparsity models and shared subspace models, as well as corresponding theories and algorithms. This framework exploits the structured information inherent in the signal, such as Block sparsity and Tree Sparsity to significantly reduce the required number of measurements while ensuring the robustness of the recovery process. La and Do developed a tree sparse signal model and a novel algorithm, Tree-based Orthogonal Matching Pursuit (TOMP), which utilizes the tree structure sparseness for signal reconstruction The results of numerical experiments show that the TOMP algorithm is significantly superior to the conventional BP and OMP methods in terms of both reconstruction quality and speed[29]. Baraniuk et al. have developed a theory of model-based compressed sensing [32]. that integrates structured sparsity models with compressed sensing, thereby proposing a novel signal recovery framework.

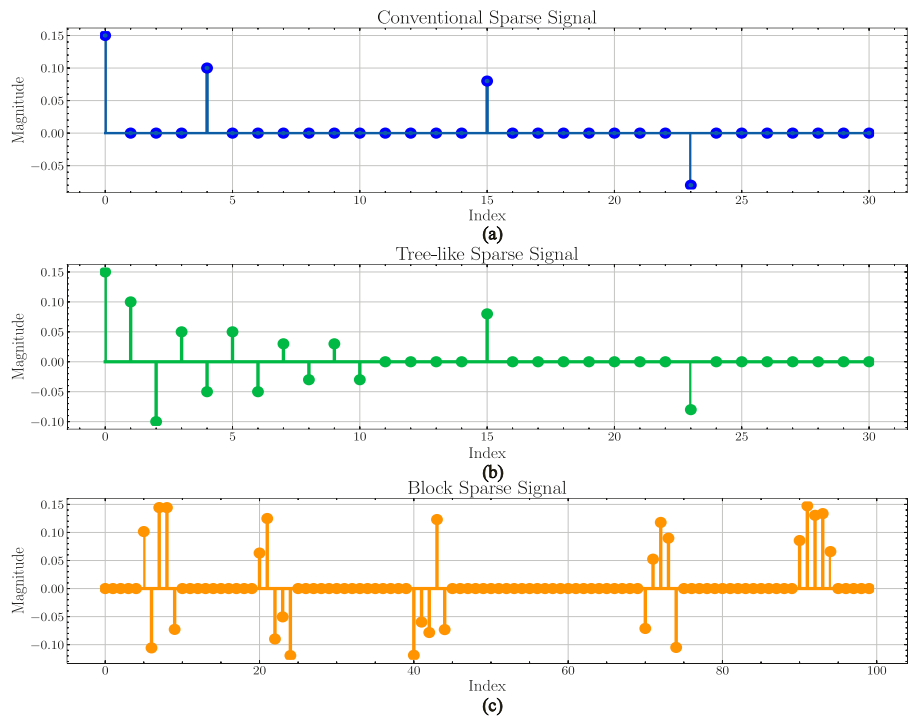

Figure 3 Standard sparseity, Tree sparsity and block sparsity

Fig.3 illustrates distinct forms of sparsity. Subfigure (a) depicts the conventional sparsity, whereas subfigures (b) and (c) illustrate two prevalent structured sparsity patterns: tree-like and block-like. The tree-like sparsity is visually misleading and can be confused with non-sparse configurations. By contrast, the block-like sparsity exhibits more pronounced characteristics.

Eldar et al. put forth a block sparse signal model and introduced the block RIP, Analogy of RIP in block sparse signals[26]. The Block RIP is a generalization of the standard RIP and provides a looser condition for sparse signal recovery.

**Definition 2:** A matrix A is said to satisfy the Block RIP with constant $\delta_B$ if, for every block sparse vector x (a vector that consists of k non-zero blocks), the following inequality holds:

$$(1 - \delta_B)||x_B||_2^2 \leq ||\Theta_B x_B||_2^2 \leq (1 + \delta_B)||x_B||_2^2 \tag{18}$$

Here, $||x_B||_2^2$ denotes the Euclidean norm of the vector $x_B$, and $||\Theta_B x_B||_2$ represents the Euclidean norm of the matrix-vector product $\Theta_B x_B$.

The block-restricted isotropic property (block-RIP) guarantees that the matrix A approximately preserves the Euclidean distance between block sparse vectors. In other words, it ensures that matrix A does not significantly distort its lengths when sparse block vectors are multiplied by the matrix. Related research shows that the conditions of the block-restricted isotropic property are more relaxed than those of the standard RIP because they fully utilize the structural information of block sparsity.

Stojnic et al. established an algorithm based on the zero-space property rather than RIP, in view of the limitations of prior knowledge of block location and size. This algorithm can use a classical measurement matrix rather than a block diagonal matrix[33]. Gao et al. demonstrate the effectiveness of a block norm minimization method for signal recovery, which can effectively recover low-block signals under the block RIP condition[34]. Cevher et al. present a novel K-C block sparse signal model for detecting K-C sparse signals in N dimensions. This model allows non-zero coefficients to be included in a maximum of C-clusters without requiring prior knowledge of the positions and sizes of these clusters[35].

These studies illustrate the potential of structured sparsity models and union subspace models for a variety of signal processing applications. Incorporating structural information about the signal can not only improve the efficiency and accuracy of signal recovery, but also shorten the required measurement length, which is particularly beneficial in resource-limited applications. It is noteworthy that although block sparse signals can be considered a special case of standard sparse signals and have received relatively limited attention, the theory of block sparse signals was developed almost simultaneously with compressed sensing. While the transformation of the sparse form of the signal into block sparsity due to SL is a difficult phenomenon to avoid, examining the aforementioned issue from the standpoint of block sparsity theory will not only make the analysis process of CSS easier, but it will also give a deeper understanding of its characteristics and new perspectives on possible methods to enhance the analysis process of CSS.

**WFs' impact on RIP and Compressed Spectrum Sensing with block sparsity**

In order to facilitate the general measurement of the WF's suppression on the SL, the following criterion is introduced. the work of Blumensath[36], when considering the effects of block RIP and sub-block sparsity terms, factors such as WF, signal subspace model, and total length must be considered. This study explores these factors and establishes a lower bound for the measurement of windowed block sparse signals. we now turn to the impact of WFs on CSS under the block sparsity model.

**WFs' impact on RIP**

**Edge Zeroing Coefficient**

In general, the SL can be mitigated to some extent by smoothing the sudden change at the boundary. Therefore, all current WFs but the rectangular WF can suppress SL to a certain extent. However, there are currently no universal criteria that can effectively measure the suppression of SL by different WFs. To address this issue, we introduce the Edge Zeroing Coefficient (EZC) as a means of evaluating the impact of WFs, which is defined as follows:

For a given WF $W(t)$ its EZC $C_Z$ can be calculated as:

$$C_Z = \int_0^{t_z} W(u)du \tag{19}$$

$C_Z$ is the integral starting from the left side of the WF, and the integral interval is small part (In our study, we use 1/20)of the total signal duration. Based on the semi-monotonicity of the WF, this simple integral term can effectively measure the smoothness of the WF at the boundary. We denote $C_Z$ as the boundary zeroing constant. Shown below are the sizes of various WFs and their corresponding EZCs, as well as the Fourier transforms of the corresponding WFs. From the figure, we can see the correspondence between the EZC and the concentration of the main lobe. The EZC allows us to effectively compare the effects of different WFs.

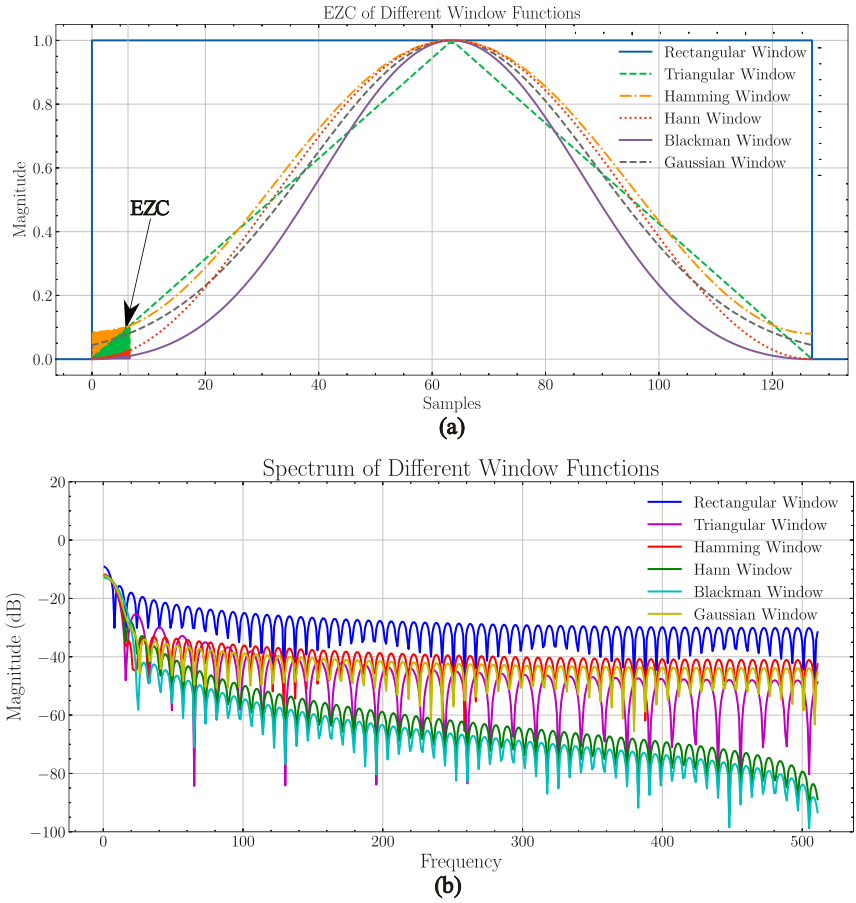

Figure 4  EZC of various wondow founction

In subfigure (a), the integral values of various WFs in the vicinity of the boundary, also known as EZC, are displayed. Experimental results indicate that the Blackman WF exhibits the minimum EZC integral value. Correspondingly, in subfigure (b), the sidelobe energy of the Blackman WF is also the lowest. By comparing and analyzing the characteristics of different WFs, it can be concluded that WFs with smaller EZC integral values tend to possess superior energy concentration capabilities.

Table I  EZC of various wondow founction

| WF | EZC | BC | ZOS | FOS | NZE |
| --- | --- | --- | --- | --- | --- |
| Rectangular | -1.31 | False | Low | Low | 0 |
| Triangular | -2.60 | True | High | Medium | -0.77 |
| Hamming | -2.36 | False | Medium | Low | -0.46 |
| Hann | -3.39 | True | High | Medium | -1.08 |
| Blackman | -3.82 | True | High | High | -1.34 |
| Gaussian | -2.52 | False | Medium | Low | -0.76 |

In Table I, 'BC' denotes the boundary continuity of the WF, while 'ZOS' and 'FOS' represent zeroth-order and first-order continuity, respectively. As discussed in Section 2.1, the higher the boundary continuity of the WF, the more significant its effect is in suppressing SL. Additionally, the 'NZE' metric reflects the proportion of nonzero elements exceeding the threshold, which shows a positive correlation with the effect of SL suppression. Table I clearly illustrates the relationship between EZC, the zeroth-order and first-order continuity of the WF, and the number of nonzero elements after transformation, which aligns with the conclusions drawn from Figure 4. Although multiple indicators need to be considered to evaluate the suppression of SL by the WF, EZC, as a quantitative indicator, provides a convenient assessment method to determine the influence of WFs on the sparsity of multi-frequency signals after Fourier transform. Furthermore, EZC is not only an effective tool to evaluate the suppression effect of WFs on SL but also useful for investigating the impact of WFs on the lower bound of the RIP of the measurement matrix.

**RIP of windowed measurement matrices：**

Before delving into the RIP for windowed measurement matrices, it is crucial to elucidate a fundamental yet often overlooked fact: RIP is inherently tied to both the sampling and reconstruction processes. This distinction becomes particularly significant when dealing with signals that are typically sparse in transform domains, such as the frequency domain. In the context of such signals, the reconstruction objective can be categorized into two main approaches:

1. Reconstruction of the sparse transformed signal (e.g., spectral analysis in audio and vibration)
2. Reconstruction of the original signal (e.g., audio and speech processing, image processing)

While these approaches may utilize the same measurement matrix, their reconstruction matrices differ substantially. This difference is critical in determining the appropriate reconstruction method and interpreting the results in various applications of compressed sensing.

Now, we examine the RIP for window measurement matrices. In conventional spectrum analysis, truncations and spectrum loss due to mismatch in the base significantly affect the accuracy of spectrum analysis. It is unfortunate that the Fourier transform is still used as a pre-sparsity method in the sensor matrix of Compressed Sensing (CS) analysis as follows.

$$\Theta_F = \Psi F \tag{20}$$

Where $\Theta_F$, $\Psi$, $F$ denotes the measurement matrix, sensing matrix and Fourier transform matrix respectively. This indicates that even in sub-Nyquist sampling again, the effects of SL cannot be circumvented.

WFs applied to a signal can be expressed in the following way:

$$\tilde{x}_W = F(W \odot x) \tag{21}$$

Where $W$, $\odot$, $F$ denotes the WF, element-wise product and Fourier transform matrix respectively.

Based on the description of the WF in Section 2, Substituting (20)(21) into (16) gives

$$y_W = \Psi F(W \odot x) + n \tag{22}$$

In (7), the WF $W$ is coupled with $x$ in the form of a vector by element-wise multiplication, rather than a matrix product with $F$. In order to achieve formal consistency and facilitate the study of the effect of the WF on the measurement matrix, we can leverage the following property of a diagonal matrix: for a diagonal matrix $D$ and a matrix or a vector $M$, their product can be expressed as follows:

$$\begin{aligned} DM &= \begin{pmatrix} d_1 & & & \\ & d_2 & & \\ & & \ddots & \\ & & & d_N \end{pmatrix} M \\ &= [d_1 m_{T1}, \cdots, d_i m_{Ti}, \cdots, d_N m_{TN}] \end{aligned} \tag{23}$$

Where $d_i, m_{Ti}$ denotes the elements on the diagonal of the diagonal matrix $D$ and the rows of transposed $M$. Given the $D_W$ is diagonal matrix with the diagonal elements corresponding WF $W$'s elements. Then with (8), (7) can be reformed as:

$$\begin{aligned} y_W &= \Psi F(W \odot x) + n \\ &= \Psi F(D_W x) + n \\ &= \Theta_F D_W x + n \\ &= \Theta_{FW} x + n \end{aligned} \tag{24}$$

To the definition of RIP, we need investigate the relationship of $\|\Theta_{FW} x\|_2$ and $\|x\|_2$, where $\Theta_{FW}$ is windowed measurement matrix. Although there are many methods for the above research, such as singular value distribution of $\Psi_W$ [37] and Johnson-Lindenstrauss (J-L) lemma [26], etc., these methods often involve complex deformation random matrix theory[38] in this context. In contrast, the 2-stability theory of Gaussian distribution can simplify this problem[39]. Therefore, we will first demonstrate the application of 2-stability of Gaussian distribution in RIP proof based on the results of [26] and [39]. Then, of course, we can extend this process to RIP of windowed CSS easily.

To the 2-stability of the Gaussian distribution: for any real numbers $x_1, x_2, \ldots, x_N$ of $x \in \mathbb{R}^d$, if $\Psi$ be a matrix of size $M \times N$ whose entries are drawn i.i.d. from $\mathcal{N}(0, 1/\sqrt{M})$.

$$Y_j = \Psi_{rj} x = \sum_{i=1}^{N} \psi_{ji} x_i, \text{then } Y \stackrel{D}{=} c\mathcal{N}(0, 1/\sqrt{M}) \tag{25}$$

Where $\Psi_{rj}$ denotes j rows of $\Psi$, and $c = (x_1^2 + \cdots + x_N^2)^{1/2}$,

As a result, if we interpret each of the k projection lengths as a coordinate in $\mathbb{R}^k$, then the squared length of the resulting vector follows the Gamma distribution for which strong concentration bounds are readily available.

Then, for $\Psi_{rj}$ denotes the $j^{th}$ row of $\Psi$ and all $x \in \mathbb{R}^d$, Given

$$f(x) = (\Psi_{r1} \cdot x, \ldots, \Psi_{rM} \cdot x).$$

Let us start by computing $\mathbf{E}(\|f(x)\|^2)$ for an arbitrary vector $x \in \mathbb{R}^N$. $\{Q_j\}_{j=1}^{k}$ is defined as

$$Q_j = \Psi_{rj} \cdot x \tag{26}$$

Then

$$E(Q_j) = E\left(\sum_{i=1}^{N} \psi_{ji} x_i\right) = \frac{1}{\sqrt{N}} \sum_{i=1}^{N} x_i E(\psi_{ji}) = 0, \tag{27}$$

and

$$\begin{aligned}
E(Q_j^2) &= \frac{1}{N} E\left(\left(\sqrt{M} \sum_{i=1}^{N} \psi_{ji} x_i\right)^2\right) \\
&= \frac{M}{N} E\left(\sum_{i=1}^{N} (\psi_{ji} x_i)^2 + \sum_{p=1}^{N} \sum_{q=1}^{N} 2\psi_{jp} \psi_{jq} x_p x_q\right) \\
&= \frac{M}{N} \sum_{i=1}^{n} x_i^2 E(\psi_{ji}^2) + \frac{1}{N} \sum_{p=1}^{N} \sum_{q=1}^{N} 2\alpha_p \alpha_q E(\psi_{jp}) E(\psi_{jq}) \\
&= \frac{M}{N} \|x\|^2,
\end{aligned} \tag{28}$$

Note that to get (27) and (28) we only used that $\psi_{ji}$ are independent, $E(\psi_{ji}) = 0$ and $\text{Var}(\psi_{ji}) = 1$. Using (28) we get

$$\mathbf{E}(\|f(x)\|^2) = \frac{N}{M} \times \sum_{i=1}^{M} E(Q_j^2) = \|x\|^2. \tag{29}$$

And then with lemma 5.1 in [26], For all $x \in \mathbb{R}^N$

$$(1-\delta)\|x\|_{\ell_2^N}\leq\|\Phi(\omega)x\|_{\ell_2^n}\leq(1+\delta)\|x\|_{\ell_2^N}. \tag{30}$$

with probability $P_s$

$$P_s \geq 1 - 2(12/\delta)^k e^{-c_0(\delta/2)n}.$$

The above are all proofs of the behavior of random projections and their RIPs for the standard measurement matrix. For the RIP proofs for the windowed measurement matrix, referring to the proof process in (25) (27), and (28), we give the results directly.

For $\Psi_W$:

$$\begin{aligned}\Psi_W ji &= \psi_{ki}w_i \\ Y_w &\stackrel{D}{=} c_w N(0,1)\end{aligned} \tag{31}$$

Where $c_w = (w_1^2 x_1^2 + \cdots + w_n^2 x_n^2)^{1/2}$.

For $W$, $c_w$, the first mean value theorem states that

$$\sup_{t\in\mathbb{R}}|W(t)x(t)|^2 \leq \alpha' \int_{\mathbb{R}} |W(u)x(u)|^2 du = \alpha_w \|x\|_{L^2(\mathbb{R})}^2. \tag{32}$$

Obviously, there exists $c_w = \alpha_w c$, where $\alpha_w = \epsilon_w(w_1^2 + \cdots + w_n^2)^{\frac{1}{2}}, 0 < \epsilon_w < 1/n$. Here, we call $\alpha_w$ the Window Scaling Coefficient (WSC).

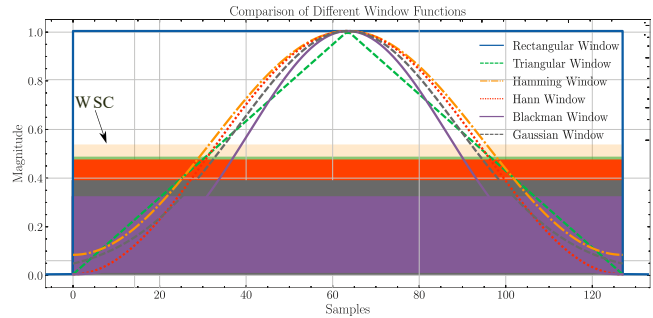

Figure 5　WSC of various WFs

Figure 5 presents a comparison of the Window Spectral Concentration (WSC) for various WFs in signal processing, allowing us to visually assess the performance of each WF on the WSC metric. Consistent with the trend of the Equivalent Zero-filled Length (EZC) metric, the Blackman WF achieves the lowest WSC value, indicating the most significant degradation in the corresponding measurement matrix's RIP . This observation confirms our previous conclusion that EZC can, to some extent, reflect the influence of WFs on the RIP of the measurement matrix. Consequently, this finding provides an important reference for selecting an appropriate WF during the process of signal reconstruction.

Table II   The Windowed RIC for Different WFs

| WF | WSC | ULref | ULexp | UBref | UBexp |
|---|---|---|---|---|---|
| Rectangular | 1 | 0.7 | 1.3 | 0.6227 | 1.2282 |
| Triangular | 0.5 | 0.35 | 0.65 | 0.3801 | 0.7257 |
| Hamming | 0.5395 | 0.3776 | 0.7019 | 0.3104 | 0.6727 |
| Hann | 0.4995 | 0.3496 | 0.6495 | 0.385 | 0.7437 |
| Blackman | 0.4196 | 0.2937 | 0.5455 | 0.2244 | 0.4512 |
| Gaussian | 0.4946 | 0.3463 | 0.6434 | 0.2706 | 0.6544 |

Table 2 presents the theoretical lower bound (ULref) and upper bound (UBref) of the windowed RIP for various WFs, with these parameters based on the normalized results of Equation (35). Additionally, the table includes experimentally measured lower (ULexp) and upper (UBexp) bounds. Comparisons show consistency between theoretical and experimental values.

Therefore, referring to (9), (12), (13), and $\alpha_w$:

$$E(||f_w(x)||^2) = \alpha_w \times E(||f(x)||^2) = \alpha_w^2 ||x||^2 \tag{33}$$

Therefore, for the RIP of $\Psi_W$:

$$(1 - \epsilon^2)\alpha_w^2 ||x||_2^2 \leq ||\Phi D_W x||_2^2 \leq (1 + \epsilon^2)\alpha_w^2 ||x||_2^2 \tag{34}$$

$$(1 - \epsilon_{WL}^2)||x||_2^2 \leq ||\Phi D_W x||_2^2 \leq (1 + \epsilon_{WU}^2)||x||_2^2 \tag{35}$$

Here $\epsilon_{WL}, \epsilon_{WU}$ are the upper and lower bounds of the corresponding windowed measurement matrices, and it is obvious that the upper and lower bounds are not symmetrical about 1 but about $\alpha_w$. It should be noted that although there is relevant literature that states that the form of the RIP can be expressed as follows

$$\beta_1 ||x||_2^2 \leqslant || \Psi x ||_2^2 \leqslant \beta_2 ||x||_2^2, \beta_2 > \beta_1 > 0$$

And it can be done to multiply $\Psi$ by $\sqrt{2/(\beta_1 + \beta_2)}$, so that $\delta_K = (\beta_2 - \beta_1)/(\beta_2 + \beta_1)$, which has no effect on the standard measurement matrix. However, after the introduction of $D_W$ when multiplied by $\sqrt{2/(\beta_1 + \beta_2)}$, $W$ deviates from the default value, generating a new WF $W\sqrt{2/(\beta_1 + \beta_2)}$, thus changing the sparsity of the components. Therefore, this theory is not applicable to Windowed CSS.

**EZC and WSC 's effect on RIP**

In this section, we will elaborate on the conclusion of EZC at the end of the first subsection in conjunction with WSC. The role of EZC is not only to measure the suppression effect of different WFs on SL, but more importantly, it also measures the impact of introducing WFs on the RIP of the measurement matrix.

Since it is a random matrix that is multiplied by the diagonal matrix of the WF, a detailed study of its singular value distribution requires the introduction of a complex deformation T random matrix theory. Fortunately, due to the special properties of the diagonal matrix with some coefficients close to 0, the relevant operations are greatly simplified and we can avoid getting bogged down in the discussion of deformed random matrices[38]

a) The degree to which a minimum singular value approaches zero determines the lower bound of the window RIP.

b) The other is the number of small singular values close to zero. By increasing the size of the zero submatrix, the null space becomes correspondingly larger and the probability that the measured vector is in the null space increases. This leads to an increase in the probability of zeroing after projection.

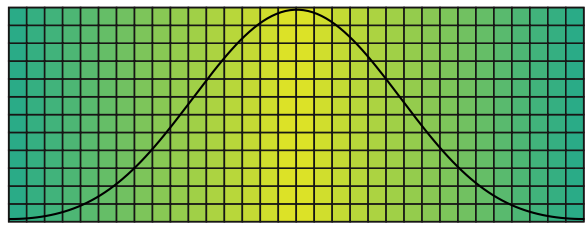

Figure 6　Boundary zeroing distribution of windowed measurement matrix

Figure 6 vividly shows the element distribution of the windowed measurement matrix, where different colors represent the amplitudes of different elements. Due to the bounded nature of the original measurement matrix, elements at the boundaries become very close to zero after windowing, leading to the possibility that some elements with close proximity might be mapped to the same value during sampling, as well as numerical instability in the reverse mapping during the reconstruction process.

Therefore, the following conclusions can be drawn from the introduction of EZC: The suppression of SL by the WF and the RIP degradation of the measurement matrix occur simultaneously. In other words, setting the bound to zero can simultaneously cause SL and deterioration of the measurement performance of the measurement matrix.

Comparing the forms of $C_Z$ and $\alpha_w$, the following conclusions can be drawn: Both $C_Z$ and $\alpha_w$ measure the effect of the WF on the measurement matrix. $C_Z$ mainly intuitively shows the effect of the WF on the lower RIP boundary from the boundary zeroing effect, while $\alpha_w$ generally characterizes the effect on the central RIP distribution. Therefore, if α is too close to 0, sparse vectors with a small distance will not be separated. The $C_Z$ and $\alpha_w$ are mutually reinforcing. The interplay between $C_Z$ and $\alpha_w$ reveals an important consideration in windowed compressed sensing: While the introduction of a WF theoretically allows the measurement matrix to map different signals to distinct observations, it may cause significant changes in the distances between signals in the observation domain. In some cases, these distances may become very close to zero, leading to reconstruction difficulties, This phenomenon underscores the importance of carefully selecting and analyzing WFs in compressed sensing applications to balance their benefits with potential challenges in signal reconstruction.

When examining the relationship between EZC and WSC, some readers might question whether one of these two entities are redundant, given that they yield similar evaluation outcomes. However, it's important to note that EZC and WSC approaches to WFs are based on distinct methodologies. EZC, as a tool for evaluating WFs, primarily focuses on the properties of WFs at their boundaries. It provides insight into how the WF behaves at its edges, which is crucial for understanding its impact on SL. In contrast, WSC is derived from stability theory and examines the overall change in the projection behavior of the measurement matrix after windowing. This approach provides a more comprehensive view of how the WF affects the entire measurement process in compressed sensing.

Given these differences, it becomes clear that EZC and WSC are complementary rather than redundant. They offer different perspectives on WF analysis, with EZC providing localized boundary information and WSC offering a global assessment of stability. Together, they provide a more comprehensive understanding of WFs and their effects on compressed sensing systems.

**Block sparse CSS model and Sample bound**

**Theory model**

As mentioned in the Introduction section, SL can be effectively suppressed by selecting an appropriate WF or significantly increasing the sample length, however, it is almost impossible to completely eliminate SL. Signal processing techniques based on the discrete Fourier transform almost always involve SL. This means that CSS techniques naturally face the basis mismatch . Fortunately, thanks to the work of Eldar [11], Lu [12]and Baraniuk [32]et al. on block-sparse signal theory, we can leverage prior knowledge of block-sparse signals to effectively improve sampling efficiency and reconstruction accuracy. Additionally, we can build on our previous research in section 2.1 to further optimize the performance of the algorithm and fully assess its potential risks.

Our research is based on model-based compressed sensing, a method for recovering a structured, sparse signal from a compressed measurement. More specifically, it is inspired by Cevher 's work with $(K, C)$ sparse mode, This model allows main coefficients of sparse signals to appear in up to C-clusters of unknown size and position. Being different from the block sparsity model in [40],[41], the (K,C) model does not require the position and size of the coefficient clusters to be known in advance, and thus extends the block sparsity model of Eldar and Baraniuk.

The (K,C) model is defined as follows:

Definition 3: The (K, C)-sparse signal model $\mathcal{M}_{(K,C)}$ is defined as

$$\mathcal{M}_{(K_S,C)} = \left\{ x \in R^{N+2} \middle| \sum_{i=1}^{2C+1} \beta_i = N_S + 2, \sum_{i=1}^{C} \beta_{2i} = K_S \right\}. \tag{36}$$

(3)

where $\beta_i$ $\beta_{2i}$ corresponds to the size of the zero sub-blocks and the non-zero sub-blocks, respectively.

By definition 3 and the proof in [], the model's subspace count is

$$m_{(K_S,C)} = \binom{N_S + 1 - K_S}{C}\binom{K_S - 1}{C - 1} \tag{37}$$

Where $N_S$ is the signal length, $K_S$ is the total number of non-zero elements, and C is the number of groups.

Based on the work of Blumensath[36] for model-based CS, the number of measurements M necessary for a subgaussian CS matrix to have the $\mathcal{M}_K$-RIP with constant $\delta_{\mathcal{M}_K}$ and with probability $1 - e^{-t}$ to be

$$M \geq \frac{2}{c\delta_{\mathcal{M}_K}}\left(\ln(2m_{(K_S,C)}) + K_S\ln\frac{12}{\delta_{\mathcal{M}_K}} + t\right). \tag{38}$$

Where $\delta_{\mathcal{M}_K}$ is the RIC of model-based CS, c is a constant depending on the probability distribution of the entries in.

For conventional CS, This bound can be used to recover the result by substituting $m_K = \binom{N_S}{K_S} \approx (N_S e/K_S)^{K_S}$.

Plugging (37) into (38), we obtain the sampling bound for M(K,C):

$$M = \mathcal{O}\left(K_S + C\log\frac{N_S}{C}\right). \tag{39}$$

From a block sparsity perspective, multi-frequency signals and the Cevher's (K,C) model share fundamental structural similarities in their block-wise signal representation. However, based on preliminary theoretical analysis, the block sparse structure of multi-frequency signals is distinctively influenced by three key factors:

(1) signal length and WF, which affects both the resolution and size of individual blocks;

(2) component frequency distribution, which determines the spacing and overlap of frequency blocks;

(3) noise floor, which impacts the detectability and effective sparsity of frequency components.

These factors collectively shape the dynamic nature of block sparsity in CSS multi-frequency signals, distinguishing them from the more constrained structure of the Cevher's model.

$$m_{(K_S,C)} = \binom{N_S + 1 - K_S}{C}\binom{K_S - 1}{C - 1}$$

In the following study, we explore how these three elements affect the sample bound of the sparse model of multi-frequency signal blocks.

**Signal length and WF**

In analyzing multi-frequency signals, while the number of signal components $K_c$ corresponds to the parameter C in the Cevher's (K,C) model and remains constant, both the total number of non-zero coefficients and the size $B_j$ of each component's frequency-domain block are dependent on the sampling frequency $f_s$ and signal length $N$. This relationship can be expressed as:

$$B_j \propto C \, \beta_j(N_S) \cdot \frac{1}{f_s} \tag{43}$$

where $C$ is constant value with the signal's property, $\beta_j$ is inversely proportional to the signal length $N_S$. the number of possible subspaces $m_{(K,C)}$ can be expressed as:

$$m_{(K_c,C)} = \binom{N_S + 1 - \sum_{j=1}^{K_c} B_j(N_S)}{K_c} \binom{\sum_{j=1}^{K_c} B_j(N_S) - 1}{K_c - 1}. \tag{44}$$

The signal length $N_S$ plays a crucial role in determining the block structure of a signal, influencing both the size of individual component blocks and the total number of non-zero coefficients. While increasing the sampling length can reduce $\sum_1^{K_c} B_i(N)$, it also increases the value of $m_{(K,C)}$. Additionally, longer sampling lengths lead to larger matrix dimensions, which can introduce computational challenges.

In contrast, windowing functions can provide a more economical approach to reducing signal sampling at shorter sample lengths. The WF's subspaces $m_{(K,C)}$ can be expressed as:

$$m_{(K,C)} = \binom{N_S + 1 - \sum_1^{K_c} B_j(W)}{K_c} \binom{\sum_1^{K_c} B_j(W) - 1}{K_c - 1}. \tag{45}$$

For spectrum reconstruction, the choice can be totally based on the EZC criterion. However, when reconstructing the original signal, as previous research has shown, excessively low EZC values may lead to instability in the reconstruction process. Hence WFs are generally more economical than increasing signal length, it's crucial to consider the potential negative impacts of very small EZC and WSC when reconstructing the original signal.

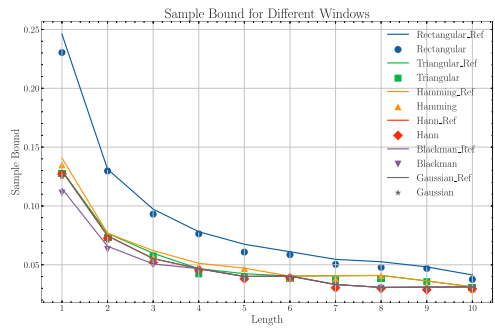

Figure 9　Sample bound of different WFs

Figure 9 shows that different WFs lead to different block sizes, which in turn affect the subspace count and ultimately the sampling bounds. It is clearly observed from the figure that the signals with no window or rectangular window achieve a higher perfect reconstruction sampling bound compared to those with other WFs. The Blackman WF exhibits the lowest sampling bound. These discussions are based on the perspective of spectral reconstruction, but as previously mentioned, the influences of EZC and WSC should be carefully considered when reconstructing the original signal.

**Component frequency distribution**

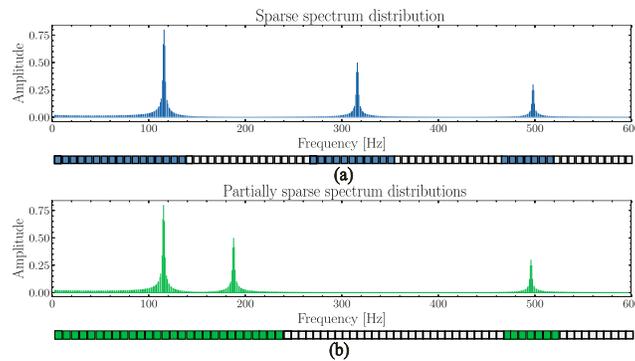

Figure 7 shows the sparse structures corresponding to different frequency distributions.

Figure 7 demonstrates that components that are farther apart in the frequency domain are typically associated with multiple smaller blocks, leading to a greater number of subspaces, indicating a higher frequency of occurrence. Conversely, components that are closer together in the frequency domain tend to form fewer but larger blocks, corresponding to a reduced number of subspaces, suggesting a lower frequency of occurrence.

The spatial distribution of frequency components fundamentally influences the block structure in the frequency domain. When frequency components are closely spaced, their individual spectral representations may overlap and coalesce into larger consolidated blocks, while maintaining the total number of non-zero coefficients. This frequency-dependent block merging phenomenon significantly affects the subspace distribution of the signal.

For a signal with $K_c$ components, the probability distribution of different sparse structures varies according to the frequency spacing. Specifically, when these $K_c$ components are partitioned into C groups due to frequency distribution. the number of possible subspaces $m_{(K,C)}$ can be expressed as:

$$m_{(K_c,C)} = \binom{N+1-\sum_{j=1}^{K_c} B_j}{C}\binom{K_c-1}{C-1}(\prod_{i=1}^{K_c} i). \tag{40}$$

Where $C = 1,\cdots,K_c$.

And the corresponding probability can be expressed as:

$$p_i = \frac{m_{(K_c,C_i)}}{\sum_{j=1}^{K_c} m_{(K_c,C_j)}}. \tag{41}$$

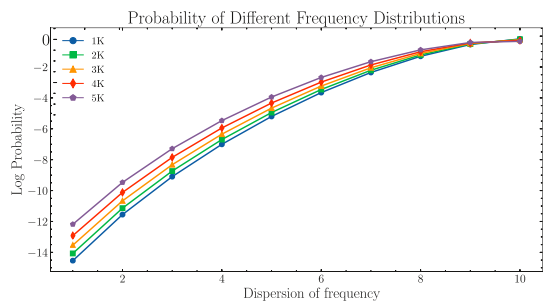

Figure 8 Probability of different frequency distributions

Figure 8 shows the probability distribution diagram under different distributions. The graph depicts the probability changes for different frequency distributions as the sub-block size increases, under given component numbers and signal length. The results indicate that under specific sub-block sizes, sample lengths, and component counts, a wide frequency band with a finite number of possible components tends to generate a sparse frequency structure. A similar conclusion can also be drawn from the perspective of Shannon's information entropy [42].

For block-sparse signals with low noise floor and large distance between components, they can be further simplified as ultrasparse-

structured signals. The number of subspaces can be simplified as follows:

$$m_{(K_c,C)} = \binom{N_S + 1 - K_c B_s}{K_c}. \tag{42}$$

From above, it can be seen that when the noise floor is relatively low and the signal is sparsely distributed, the number of subspaces corresponding to the signal is significantly reduced. This means that when spectrum sensing ultra-wideband sparse signals using block sparsity-based CSS, higher reconstruction accuracy can be achieved with a smaller number of samples compared to conventional CSS.

**Noise floor**

The noise floor significantly influences the block sparse structure of frequency-domain signals by determining the detectability of spectral components above the noise floor. Specifically, noise floor affects the size and number of sub-blocks, and consequently modifying the total number of non-zero coefficients and block dimensions.

$$m_{(K_c,C)} = \binom{N_s + 1 - \sum_{i=1}^{K_c} B_i(N_s)}{K_c} \binom{\sum_{i=1}^{K_c} B_i(N_s) - 1}{K_c}. \tag{46}$$

A critical observation is that while increased noise floor elevates the average noise level, it paradoxically appears to enhance signal sparsity under identifiable conditions. This occurs as fewer spectral components exceed the detection threshold, reducing the number of observable non-zero elements and associated subspaces. However, this apparent increase in sparsity is misleading, as it comes at the cost of signal fidelity and represents a degradation rather than an improvement in signal representation.

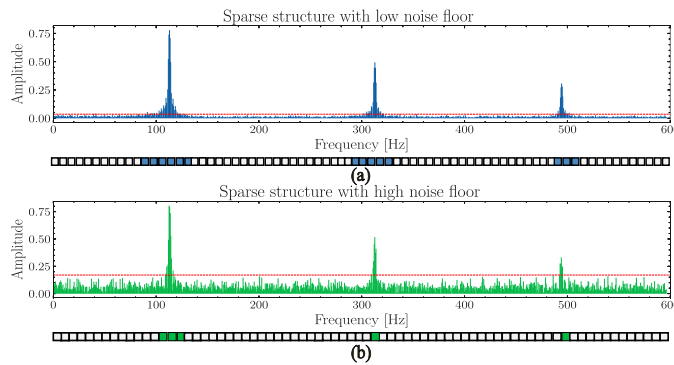

Figure 10 depicts different base noise levels corresponding to sparse structures.

**Discussion**

Our study addresses the research gap on the impact of SL on CSS by investigating the effect of WFs on signal sparsity and measurement matrices, proposing a sparse signal model accounting for SL, and studying sampling lower bounds via subspace counting. We introduce two novel metrics: EZC and WSC. EZC quantifies the SL suppression effect of various WFs and evaluates their influence on the deterioration of RIP. WSC clarifies the quantitative relationship between windowed and standard RIP, revealing the compressing effect on the distance between projected vector pairs. Our findings demonstrate that while WFs reduce SL, excessively small EZC and WSC values can negatively affect RIP quality and lead to numerical instability during reconstruction.

The introduction of WFs in CSS presents both advantages and challenges. On the positive side, WFs reduce the approximate sparsity of the signal, potentially improving reconstruction accuracy. However, they also cause RIP degradation due to the boundary zeroing effect, which can impact the stability of signal reconstruction. Our analysis demonstrates that careful selection of WFs is crucial to balance SL suppression with maintaining RIP quality. We propose a block-sparse approach to CSS, which enables more precise compression and reconstruction, particularly for low noise floor and super-sparse signals. This approach accounts for the band-like transitions of non-zero elements caused by unavoidable SL, offering a more realistic model of signal structure in CSS applications.

Our research provides a comprehensive theoretical framework for improving signal reconstruction quality in CSS. However, there are limitations to our study that should be addressed in future research. First, our focus has primarily been on random matrices, and the effects of windowing on structured measurement matrices require further investigation. Additionally, while we have concentrated on multi-frequency signals, the application of our findings to other signal types, such as modulated signals, needs to be explored. Future work should also consider developing new WFs specifically designed for CSS applications, taking into account the unique challenges and requirements of compressed sensing techniques. Despite these limitations, our study offers valuable insights into the interplay between WFs, SL, and compressed sensing performance, paving the way for more efficient and accurate spectrum sensing techniques in various applications.

**Conclusion**

This study addresses the research gap on SL in CSS by introducing two novel metrics: EZC and WSC. These metrics, based on WFs for SL suppression and the structural characteristics of windowed measurement matrices, provide a comprehensive evaluation of WFs' impact on CSS. Our findings demonstrate that while WFs effectively reduce SL, excessively small EZC and WSC values can negatively affect the RIP quality and lead to numerical instability during signal reconstruction. This research highlights the delicate balance required when applying WFs in CSS, emphasizing the need for careful consideration of their effects on both SL suppression and measurement matrix stability.

To address the inevitability of SL in practical applications, we leverage the concept of block sparsity to construct a signal model that accounts for various noise floor. By deriving the corresponding sampling thresholds for this model, we provide a framework for optimizing CSS performance in the presence of SL. Our analysis reveals that for high SNR and super-sparse signals, the required sample bound can be significantly lower than that of conventional CSS approaches. This block-sparse approach enables more precise compression and reconstruction, particularly in challenging scenarios where traditional CSS methods may struggle due to increased non-zero elements caused by SL.

Future research should focus on three key areas to further advance the field of CSS. First, there is a need to explore new WFs specifically designed for CSS applications, as current research primarily utilizes WFs developed for traditional spectrum sensing. Second, while our study concentrates on the effects of windowing on random matrices, further investigation into other types of measurement matrices, such as structured matrices, is necessary to broaden the applicability of our findings. Finally, although this research primarily addresses multi-frequency signals, extending the application to other signal types, such as modulated signals, would significantly enhance the versatility of CSS techniques. By pursuing these avenues, researchers can build upon this comprehensive theoretical framework to further improve signal reconstruction quality and efficiency in various CSS applications.

**Acknowledgments**